\documentclass[aps,prl,twocolumn,superscriptaddress,showpacs,floatfix]{revtex4-1}

\usepackage{hyperref}
\usepackage{color}
\usepackage{graphicx}	% Include figure filed
\graphicspath{%
  {./},%
}

\usepackage{amstext}
\usepackage{array}

\usepackage{xspace}	% Include xspace

\begin{document}

\title{Non-hydrodynamic transport in trapped unitary Fermi gases}

\author{Jasmine Brewer and Paul Romatschke} 
\affiliation{University of Colorado at Boulder}
\date{\today}

\begin{abstract}
Many strongly coupled fluids are known to share similar hydrodynamic transport properties. In this work we argue that this similarity could extend beyond hydrodynamics to transient dynamics through the presence of non-hydrodynamic modes. We review non-hydrodynamic modes in kinetic theory and gauge/gravity duality and discuss their signatures in trapped Fermi gases close to unitarity. Reanalyzing previously published experimental data, we find hints of non-hydrodynamic modes in cold Fermi gases in two and three dimensions.
\end{abstract}

\maketitle

%\subsection{Introduction}

In the past decade, precision experiments of ultracold quantum gases at unitarity have enabled the study of transport phenomena in strongly coupled systems \cite{2004PhRvA..70e1401K,2007PhRvL..98q0401J,2008PhRvA..78e3609R,Cao:2010wa,Vogt:2011np,2012Sci...335..563K}. Usually, transport phenomena are encoded in hydrodynamic transport coefficients, such as the speed of sound, shear and bulk viscosities, heat conductivities, spin diffusion coefficients, etc. Presently, the speed of sound has been measured in a three dimensional unitary Fermi gas \cite{2007PhRvL..98q0401J,2012Sci...335..563K}, and the shear viscosity has been constrained for both three dimensional and two dimensional Fermi gases \cite{Cao:2010wa,Vogt:2011np}. The extremely low values of shear viscosity (when expressed in units of entropy) in particular suggest similarities between cold Fermi gases close to unitarity and very different systems such as hot quark gluon plasmas \cite{Luzum:2008cw,Heinz:2013th}, high-temperature superconductors \cite{Rameau:2014gma} and strongly coupled fluids described by black holes via the AdS/CFT conjecture \cite{Policastro:2001yc}, all of which share similar transport behavior. This apparent similarity in otherwise completely different physical systems suggests that these systems could be part of a broader class of so-called strongly interacting quantum fluids (SIQFs). 

It is conceivable that SIQFs share other properties besides their similar (hydrodynamic) transport behavior. This could be important because it could imply that it is possible to learn about one example of SIQFs (say high-temperature superconductors) through studying a different SIQF for which a particular trait is more easily accessible. In the present study we will investigate ultracold Fermi gases close to unitarity and argue that they exhibit properties similar to black hole SIQFs.

One property that is quite remarkable about black hole SIQFs is that they do not seem to possess a description in terms of weakly coupled quasiparticles \cite{CasalderreySolana:2011us}. Instead, black holes can be characterized in terms of their ring-down spectrum, similar to a glass struck (lightly) with a fork \cite{Kokkotas:1999bd}. Some of these quasinormal modes can be recognized to be the well-known hydrodynamic modes, i.e. sound and shear excitations. Others do not have an equivalent in (Navier-Stokes) hydrodynamics, and are thus non-hydrodynamic, but nevertheless affect transport properties (particularly on short time scales). 

If properties of SIQFs are universal, one would expect the presence of non-hydrodynamic  modes in cold Fermi gases close to unitarity. This provides the motivation for searching for non-hydrodynamic modes in cold Fermi gases, both theoretically and experimentally.

\paragraph{\bf Transport in hydrodynamics and beyond:}

In order to understand the properties of non-hydrodynamic transport, let us briefly recall the properties of transport within hydrodynamics. To be specific, let us treat the case of a single component uncharged fluid in $D$ spatial dimensions described by the Navier-Stokes equation (which clearly will not be applicable to Fermi gases below the superfluid phase transition temperature). The fluid is then characterized by the mass density $\rho$, the fluid velocity ${\bf u}$, the pressure $P$ and the temperature $T$. Assuming hydrodynamic transport to be dominated by the shear viscosity coefficient $\eta$, we may set the bulk viscosity and heat conductivity to zero (this is a good assumption for Fermi gases close to unitarity, \cite{Schafer:2007pr}) . Finally allowing for a force term ${\bf F}$ (e.g. through a trapping potential) the fluid must obey the equations of motion
\begin{eqnarray}
\label{eq:pmot}
\partial_t \rho+\partial_i \left(\rho u_i\right)&=&0\,,\nonumber\\
\partial_t \left(\rho u_i\right)+\partial_j \left(\rho u_i u_j+P \delta_{ij}+\pi_{\ij}\right)&=&\rho F_i\,,\nonumber\\
\partial_t {\cal E}+\partial_j\left[u_j \left({\cal E}+P\right)+\pi_{ij}u_i\right]&=&\rho F\cdot u\,,\nonumber\\
-\eta \sigma_{ij} &=&\pi_{ij}\,,
\end{eqnarray}
with ${\cal E}=\frac{\rho {\bf u}^2}{2}+\frac{D}{2} P$ and $\sigma_{ij}=\left(\partial_i u_j+\partial_j u_i-\frac{2}{D}\delta_{ij}\partial \cdot u\right)$.
To close the system of equations, we adopt an ideal equation of state implying $P=c_s^2(T) \rho$, with $c_s^2(T)$ the temperature-dependent speed of sound squared (we assume an isothermal system). Additionally, we assume $\frac{\eta}{P}={\rm const}$ for simplicity. Considering small perturbations around some (possibly space-dependent) equilibrium configuration we have $\rho=\rho_0({\bf x})+\delta \rho(t,{\bf x})$, ${\bf u}=\delta {\bf u}(t,{\bf x})$, $c_s^2=c_0^2+\delta c^2(t)$. Hydrostatic equilibrium requires $\rho_0 {\bf F}=\nabla P_0$. 
%For small perturbations, the equations of motion then become
%\begin{eqnarray}
%\label{eq:pmot}
%\partial_t \frac{\delta \rho}{\rho_0}+\nabla\cdot \delta {\bf u}+\delta {\bf u}\cdot
%\nabla \log{\rho_0}&=&0\,,\nonumber\\
%\partial_t \delta u_i+c_0^2\partial_i \frac{\delta \rho}{\rho_0}+\delta c^2 \partial_i \log %\rho_0&=&\frac{-\partial_j \delta \pi_{ij}}{\rho_0}\,,\nonumber\\
%\partial_t \delta c^2 &=&-\frac{2}{D} \nabla \cdot \delta {\bf u}\,,\nonumber\\
%-\frac{\eta c_0^2}{P_0} \left(\delta \sigma_{ij}\partial_j \log \rho_0+\partial_j \delta %\sigma_{ij}\right)&=&\frac{\partial_j \delta \pi_{ij}}{\rho_0}\,.
%\end{eqnarray}
%
For perturbations around a constant density $\rho_0$, there are two solutions for the perturbations, namely the familiar hydrodynamic sound and shear modes. It is useful to characterize these modes in Fourier space, e.g. $\rho\simeq e^{-i\omega t+i{\bf k}\cdot {\bf x}}$. For the sound mode (coupling perturbations $\delta \rho, \partial \cdot {\bf u}$) one finds the dispersion relation $\omega=\pm c_0 |{\bf k}|-i \frac{\eta{\bf k}^2 c_0^2}{P_0} \left(1-\frac{1}{D}\right)$, whereas for the shear mode (perturbations {\bf u} transverse to ${\bf k}$) one finds $\omega=-i\frac{\eta {\bf k}^2c_0^2}{P_0}$. Thus, one expects to find density perturbations which for fixed ${\bf k}$ behave as
\begin{equation}
\label{eq:hydrosound}
\delta \rho_{\rm hydro}(t,{\bf x})\propto e^{\pm i c_0 |{\bf k}| t+ i{\bf k}\cdot x -\Gamma_0 t {\bf k}^2}\,,
\end{equation}
where $\Gamma_0=\frac{\eta c_0^2}{P_0}\left(1-\frac{1}{D}\right)$. Eq.~(\ref{eq:hydrosound}) was derived using hydrodynamics, so its regime of validity is that of low ${\bf k}$.

The result for the sound mode perturbation in hydrodynamics is to be contrasted with the result found for black hole SIQFs. Ref.~\cite{Kovtun:2005ev} calculated the poles of the energy-momentum tensor correlator in the sound mode channel, finding the complete set of quasinormal modes for $D=3$. For small ${\bf k}$, the spectrum can be approximated as
\begin{equation}
\label{eq:bhexi}
\omega_{h}=\pm c_0 |{\bf k}|-i \frac{\eta {\bf k}^2}{4 P_0}\frac{2}{3}\,,\quad
\omega_{nh,1}\simeq 2 \pi T (\pm 1.73-1.34 i)\,,
%\omega_{n\gg 1}=2 \pi T n (\pm 1- i)\,.
\end{equation}
and $\omega_{nh,n\gg 1}=2 \pi T n (\pm 1- i)$. The result for $\omega_h$ is just that of a hydrodynamic dispersion relation for relativistic fluids in $D=3$ (note that for Ref.~\cite{Kovtun:2005ev} $\frac{\eta}{4 P_0}=\frac{1}{4\pi T}$ and hence $\omega_{nh,n\gg 1}= \frac{2 P_0 n}{\eta}(\pm 1-i)$ is also a consistent interpretation). The presence of the non-hydrodynamic modes $\omega_{nh}$, however, implies that density perturbations for fixed $|{\bf k}|\ll 1$ behave as
\begin{equation}
\label{eq:qnm}
\delta \rho_{BH}(t,{\bf x})\propto e^{\pm i c_0 |{\bf k}|t +i {\bf k}\cdot {\bf x}-\Gamma_{h} t }+\sum_{n=1}^\infty a_{nh,n} e^{\pm i {\rm Re} \omega_{n} t -\Gamma_{n} t}\,,
\end{equation}
with $\Gamma=-{\rm Im} \omega$ and amplitudes $a_{nh,n>0}$ for the non-hydrodynamic modes. Note that an infinite tower of quasinormal modes has also been found in non-relativistic systems for black hole SIQFs, cf.~\cite{Schaefer:2014aia}.
Also, non-hydrodynamic modes have been discussed before in the context of condensed matter systems in Ref.~\cite{cmnonhy}.

Unless the amplitudes $a_n$ are extremely small or the damping rates $\Gamma_n$ are very large, one could expect the non-hydrodynamic contributions to the evolution of density perturbations to be experimentally observable. In the following, we will study the case of cold Fermi gases close to unitarity and investigate whether they exhibit non-hydrodynamic behavior similar to Eq.~(\ref{eq:qnm}).

\paragraph{\bf Non-hydrodynamic collective modes for a trapped Fermi gas:}

The analysis of collective modes changes if the equilibrium background density $\rho_0({\bf x})$ is space-dependent, as is the case for experiments on Fermi gases in a trap.  For an idealized harmonic trapping potential with base frequency $\omega_\perp$, the background density may then be written as
\begin{equation}
\label{eq:potential}
\log \rho_0({\bf x})=-\frac{\omega_\perp^2 (x^2+y^2+\lambda^2 z^2)}{2 c_0^2}\,,
\end{equation}
where $\lambda=0$ for $D=2$ and $\lambda\ll 1$ for $D=3$ (elongated trap).
It is possible to classify all collective modes in this case by realizing that in the isothermal limit, $\delta c^2(t)$ is only a function of time, and hence the equations of motion place constraints on the space-dependence of $\nabla\cdot \delta {\bf u}$. Thus, rather than analyzing perturbations in Fourier space, it is useful to expand perturbations in terms of a power series of $x,y,$ and $z$. Restricting to inequivalent polynomials under rotations in the $x,y$ plane, this leads to the ansatz
\begin{equation}
\frac{\delta \rho}{\rho_0}=c_{00}(t)+c_{11}(t) r \cos{\phi}+c_{20}(t)r^2+c_{22}(t) r^2 \cos 2\phi+\ldots\,,
\end{equation}
with $x=r \cos\phi$, $y=r \sin\phi$.
Solving the equations of motion (\ref{eq:pmot}) then leads to the identification of three collective modes, corresponding to the time evolution of $c_{11}$, $c_{00}$ coupled with $c_{20}$, and $c_{22}$, respectively. These modes may be recognized to be the usual sloshing (``S''), breathing (``B'') and quadrupole mode (``Q''), respectively, and one finds the following dispersion relations in these channels:
\begin{eqnarray}
\label{eq:hymodes}
\frac{\omega_{S,h}}{\omega_\perp}&=&\pm1\nonumber\\
\frac{\omega_{B,h}}{\omega_\perp}&=&\pm\sqrt{2+\frac{4}{D}}-\frac{i \eta \omega_\perp}{P_0}\left(1-\frac{2}{D}\right)\,,\nonumber\\
\frac{\omega_{Q,h}}{\omega_\perp}&=&\pm\sqrt{2}-\frac{i \eta \omega_\perp}{P_0}\,.
\end{eqnarray}
%(Note that the sloshing mode also carries a damped contamination of the same frequency.)
All of these modes are driven versions of the sound mode excitations. One can explicitly verify that for an isothermal gas with trapping potential (\ref{eq:potential}) no shear mode perturbations are excited.

Unlike the non-hydrodynamic quasinormal modes encountered in Eq.~(\ref{eq:bhexi}), the hydrodynamic damping rates in Eq.~(\ref{eq:hymodes}) monotonically decrease in the ideal hydrodynamic limit $\eta\rightarrow 0$.

It is possible to try to extend the collective mode analysis for a trapped Fermi gas beyond Navier-Stokes hydrodynamics. One option to do this is within the framework of second-order hydrodynamics \cite{Baier:2007ix,Chao:2011cy}. While second-order hydrodynamics is able to correctly capture higher order gradient corrections (e.g. ${\cal O}({\bf k}^3)$ corrections in Eq.~(\ref{eq:bhexi})), its regime of validity is still $\omega \ll 1$. Thus one cannot expect second-order hydrodynamics to correctly capture non-hydrodynamic modes. Nevertheless, one may try to analyze collective modes for a Fermi gas in a trap using second-order hydrodynamics to see if non-hydrodynamic modes emerge at least on a qualitative level. This can be done easily since for linear response, second-order hydrodynamics basically replaces $\pi_{ij}=-\eta \sigma_{ij}\rightarrow\pi_{ij}+\tau_\pi \partial_t \pi_{ij}=-\eta \sigma_{ij}$, with $\tau_\pi$ the relaxation time for shear viscous stresses \cite{Baier:2007ix,Chao:2011cy}. Thus, one finds that in addition to the hydrodynamic modes in Eq.~(\ref{eq:hymodes}), there is a non-hydrodynamic mode for the quadrupole mode oscillation in $D=2,3$ and for the breathing mode oscillation in $D=3$, given by
\begin{eqnarray}
\label{eq:Bmodes}
\frac{\omega_{B,nh}}{\omega_\perp}=-\frac{i}{\tau_\pi \omega_\perp}\,,\quad
\frac{\omega_{Q,nh}}{\omega_\perp}=-\frac{i}{\tau_\pi \omega_\perp}\,.
\end{eqnarray}

Another option, seemingly distinct from second-order hydrodynamics, is to consider kinetic theory in the form of the Boltzmann equation. For kinetic theory with a simple collision term in the form of a relaxation time approximation, the collective modes of a Fermi gas in a harmonic trap have been analyzed before \cite{Brewer2015,PhysRevA.76.033610,2008PhRvA..78e3609R,Vogtthesis}. For a relaxation time $\tau_R$ in the collision term of the Boltzmann equation, one finds dispersion relations for the breathing and quadrupole modes given by
\begin{eqnarray}
\label{eq:bdp}
i \omega_Q (w_Q^2-4 \omega_\perp^2)\tau_R + (2 \omega_\perp^2-\omega_Q^2)&=&0\,,\quad D=2,3\nonumber\\
 %(w_B^2-4 \omega_\perp^2)&=&0\,,\quad D=2\,,\nonumber\\
i \omega_B (w_B^2-4 \omega_\perp^2)\tau_R+\left(\frac{10}{3}\omega_\perp^2-\omega_B^2\right)&=&0\,,\quad D=3\,,\hspace*{0.5cm}
\end{eqnarray}
and $\omega_B^2=4 \omega_\perp^2$ for $D=2$. Eqns.~(\ref{eq:bdp}) have three solutions for $\omega_Q,\omega_B$ each, two of which agree with the hydrodynamic modes Eq.~(\ref{eq:hymodes}) if  $\omega_\perp \tau_R=\frac{\eta}{P_0}\ll 1$. However, in addition to the hydrodynamic modes, Eqns.~(\ref{eq:bdp}) contain one non-hydrodynamic mode for both the quadrupole mode and breathing mode in $D=3$ (cf.~\cite{PhysRevA.82.023609,Brewer2015}). In the strong coupling limit $\tau_R\rightarrow 0$, these non-hydrodynamic modes obey the dispersion relations (\ref{eq:Bmodes}) upon setting $\tau_\pi=\tau_R$. (Note that in second-order hydrodynamics $\eta,\tau_\pi$ are two different parameters, while in kinetic theory $\tau_R$ controls both the viscosity and the viscous stress relaxation time). The fact that the kinetic theory result matches that from second-order viscous hydrodynamics is expected since kinetic theory is known to be a special case of second-order hydrodynamics in the strongly interacting $\tau_R\rightarrow 0$ limit. 
However, one furthermore finds that in kinetic theory, non-hydrodynamic modes are present for all values of $\tau_R$, describing a purely damped response in the quadrupole and $D=3$ breathing mode channel. In particular, one finds from Eqns.~(\ref{eq:bdp})
\begin{equation}
\label{eq:kBmodes}
\frac{\omega_{B,nh}}{\omega_\perp}=-\frac{5 i}{6 \tau_R \omega_\perp}\,(D=3)\,,
\ 
\frac{\omega_{Q,nh}}{\omega_\perp}=-\frac{i}{2 \tau_R \omega_\perp}\,,
\end{equation}
in the limit of weak interactions $\tau_R\rightarrow \infty$. 
Thus, while (\ref{eq:Bmodes}) is beyond the regime of validity in second-order hydrodynamics, Eq.~(\ref{eq:kBmodes}) is well within the regime of validity for kinetic theory (weak interactions, well defined quasiparticles).

\paragraph{\bf Experimental evidence for non-hydrodynamic collective modes:}

Summarizing the theoretical status of non-hydrodynamic transport, it is known that non-hydrodynamic modes exist for some SIQFs such as black hole duals \cite{Policastro:2001yc}. Furthermore, for trapped Fermi gases it is known that non-hydrodynamic collective modes are featured in descriptions beyond Navier-Stokes, but lie outside the regime of validity of these descriptions for strong interactions \cite{Schaefer:2014awa,Brewer2015}. If the interactions are weak, and quasiparticle degrees of freedom are well defined, kinetic theory also predicts the presence of non-hydrodynamic modes, which are purely damped excitations (\ref{eq:kBmodes}). 

In view of this, experimental input is crucial to hope to understand non-hydrodynamic transport for SIQFs. For this reason, we reanalyze the raw data for the collective oscillations observed in a $D=2$ gas of $^{40}K$ atoms from Refs.~\cite{Vogt:2011np,2013PhRvA..87d3612B} as well as raw data for the radial breathing mode oscillations observed in a $D=3$ gas of $^6{\rm Li}$ atoms close to unitarity from Refs.~\cite{2005PhRvL..94q0404K,2011Sci...331...58C}. To analyze the data for the time-evolution of the quadrupole (``Q'') and breathing (``B'') mode amplitudes, we use a fitting function of the form
\begin{equation}
\label{fittingf}
F(t)=a_{h} \cos (w_h t)e^{- \Gamma_h t}+a_{nh} \cos(w_{nh} t) e^{-\Gamma_{nh} t}\,,
\end{equation}
where irrelevant phase shifts and offsets have been suppressed. Minimum $\chi^2$ fits to the data of Refs.~\cite{Vogt:2011np,2013PhRvA..87d3612B,2005PhRvL..94q0404K,2011Sci...331...58C} are performed to extract frequencies $w\equiv {\rm Re}\,\omega$, damping rates $\Gamma \equiv- {\rm Im}\,\omega$ as well as the amplitudes $a$ for the hydrodynamic (``h'') and non-hydrodynamic (``nh'') modes. 

%\begin{figure*}[t]
%\centering
%\includegraphics[width=0.45\textwidth]{cmp_exp_w.eps}\hfill
%\includegraphics[width=0.45\textwidth]{cmp_exp_g.eps}
%\caption{Hydrodynamic mode frequencies and damping rates for the two-dimensional Fermi gas in a trap. Shown are results from the experiment by Vogt et al. in Ref.\cite{Vogt:2011np}, our re-analysis of the raw time-evoution data from Ref.\cite{Vogt:2011np}, as well as the analytic results from kinetic theory for an ideal gas in an harmonic trap  (see text for details) .
%\label{fig:one}}
%\end{figure*}

%\begin{figure*}[t]
%\centering
%\includegraphics[width=0.45\textwidth]{ratios.eps}\hfill
%\includegraphics[width=0.45\textwidth]{cmp_exp_g1.eps}
%\caption{Non-hydrodynamic amplitudes and damping rates for the two-dimensional Fermi gas in a %trap. Shown are results from our re-analysis of the raw time-evolution data from the experiment by %Vogt et al. in Ref.~\cite{Vogt:2011np}, as well as the analytic results from kinetic theory for an ideal %gas in an harmonic trap  (\ref{eq:bdp}) .
%\label{fig:two}}
%\end{figure*}

%\begin{figure*}[t]
%\centering
%\includegraphics[width=0.45\textwidth]{ratios_3D.eps}\hfill
%\includegraphics[width=0.45\textwidth]{cmp_exp_g1_3D.eps}
%\caption{Non-hydrodynamic amplitudes and damping rates for the three-dimensional Fermi gas in a %trap. Shown are results from our re-analysis of the raw time-evolution data from the experiment by %Kinast et al. in Ref.~\cite{2005PhRvL..94q0404K}, as well as the analytic results from kinetic theory for %an ideal gas in an harmonic trap (\ref{eq:bdp}).
%\label{fig:tev}}
%\end{figure*}

\begin{figure*}[t]
\centering
\includegraphics[width=0.45\textwidth]{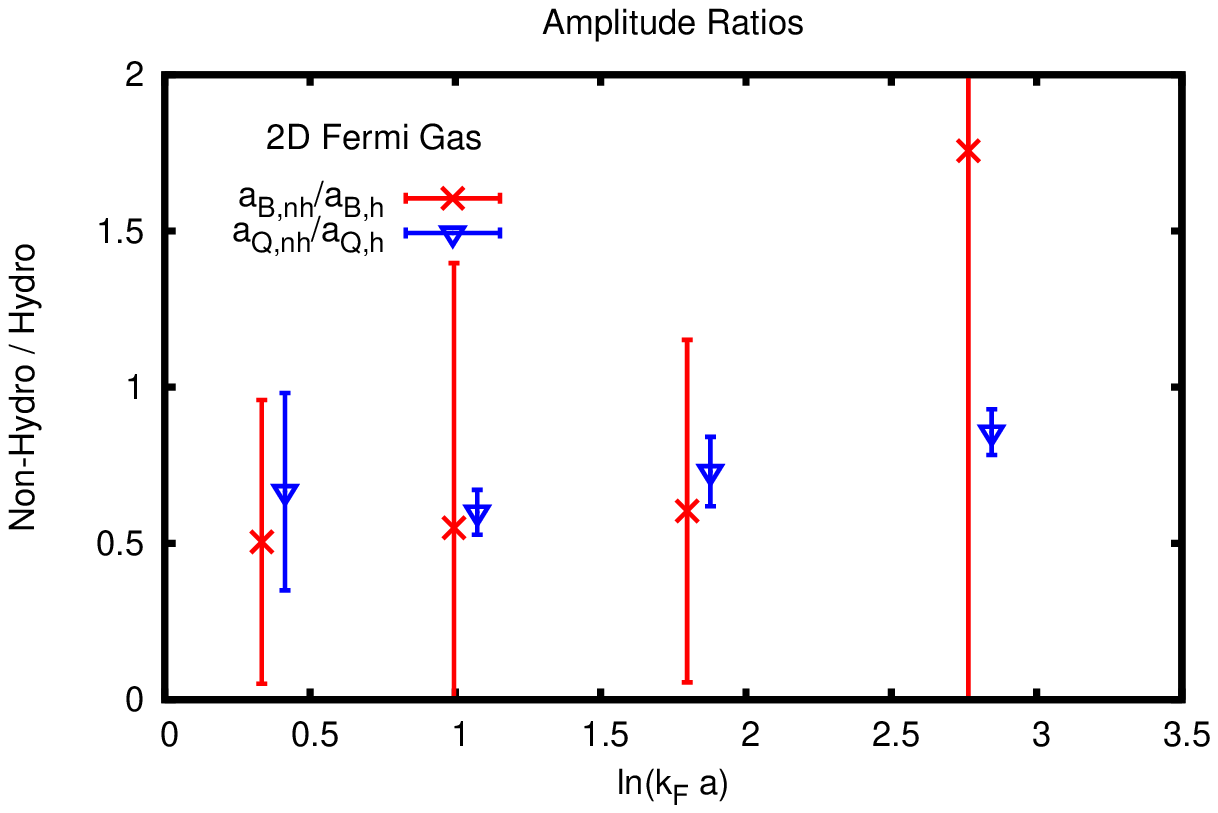}\hfill
\includegraphics[width=0.45\textwidth]{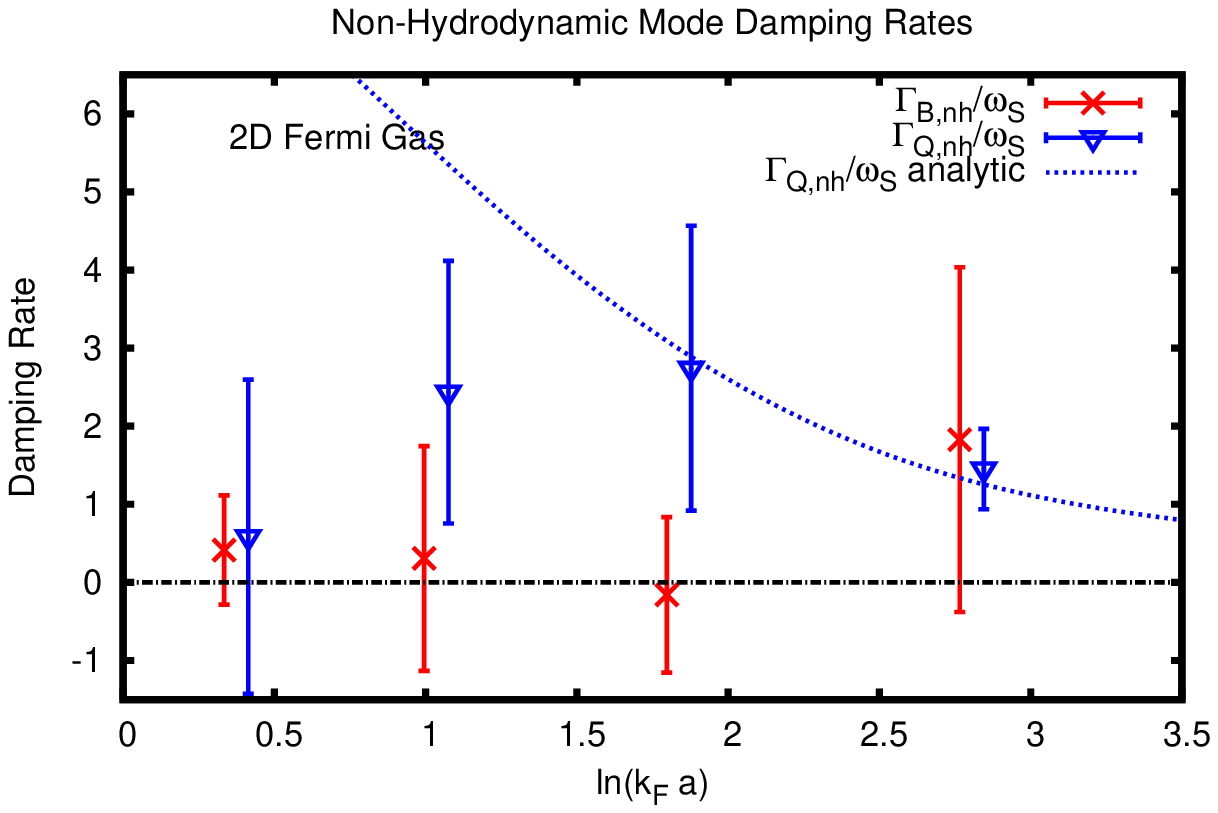}
\includegraphics[width=0.45\textwidth]{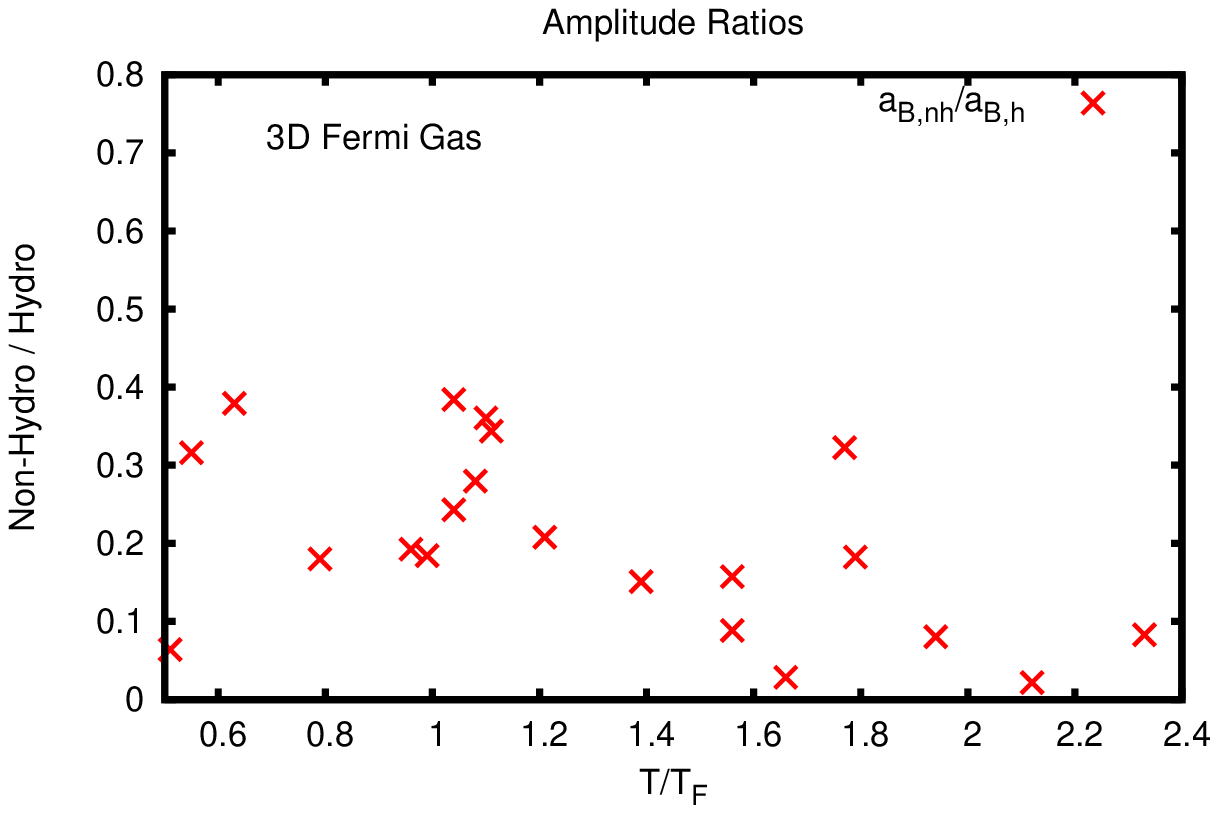}\hfill
\includegraphics[width=0.45\textwidth]{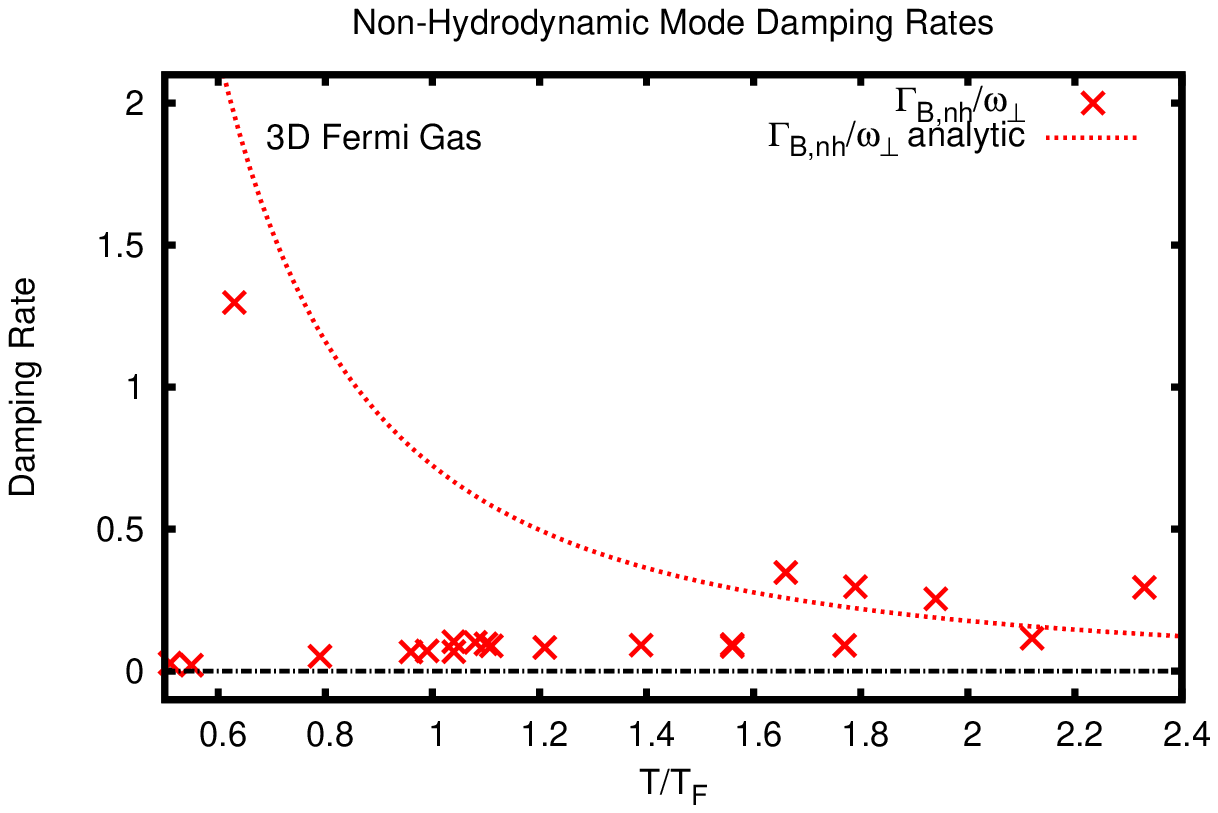}
\caption{Non-hydrodynamic amplitudes and damping rates for the Fermi gas in a trap in two dimensions (upper panels) and three dimensions (lower panels). Shown are results from our reanalysis of the raw time-evolution data from the $D=2$ experiment by Vogt et al. in Ref.~\cite{Vogt:2011np} (upper panels), from the $D=3$ experiment by Kinast et al. in Ref.~\cite{2005PhRvL..94q0404K} (lower panels), as well as the analytic results from kinetic theory with ideal gas equation ofstate in a harmonic trap  (\ref{eq:bdp}) .
\label{fig:two}}
\end{figure*}

We employed the Akaike information criterion (AIC) \cite{1974ITAC...19..716A} to avoid overfitting the raw data with too many parameters. Using the AIC values extracted for both $D=2,3$ data suggests that the currently available data is not sufficient to extract information about non-hydrodynamic frequencies. For this reason, we assume $\omega_{nh}=0$ in the following.

%\subsection{Two-dimensions}

For the case of $D=2$, we analyze raw data by Vogt et al. in Ref.~\cite{Vogt:2011np} given at $\frac{T}{T_F}=0.45$ with $T_F=\sqrt{2 N} \omega_\perp$, $\omega_\perp\sim 2\pi\times 125$ Hz for the case of $N\simeq 2000$ $^{40}K$ atoms for four different values of the interaction strength parameter $\ln(k_F a)$. At each value of $\ln(k_F a)$, six different measurements of the time evolution are available and we obtain mean values and error estimates for the fit parameters in Eq.~(\ref{fittingf}) from averaging over these six sets. For each of the $\ln(k_F a)$ values, we extract hydrodynamic mode frequencies and damping rates that are consistent with published values \cite{Vogt:2011np,2013PhRvA..87d3612B}.

%As shown in Fig.~\ref{fig:one}, our re-analysis for the hydrodynamic frequencies and damping rates from the data essentially agrees with those of Ref.~\cite{Vogt:2011np}. For comparison, analytic results for an ideal gas in an harmonic trap in kinetic theory, reducing to Eq.~(\ref{eq:hymodes}) and Eq.~(\ref{eq:kBmodes}) in the hydrodynamic and ballistic limits, respectively, are also shown in Fig.~\ref{fig:one} using the relation $\omega \tau_R=0.12 \left(1+\frac{4}{\pi^2}\ln^2(k_F a)\right)$ (cf. Ref.\cite{Brewer2015,2012PhRvA..85a3636B}). 

In Fig.~\ref{fig:two}, amplitudes of putative non-hydrodynamic modes extracted from the raw time-evolution data in Ref.~\cite{Vogt:2011np} are shown, indicating a sizeable non-hydrodynamic mode component amplitude in the quadrupole excitations. We find that for a total of 240 data-points, the AIC for the quadrupole mode decreases by 80 units when allowing for a non-hydrodynamic mode with $a_{nh}\neq 0$ to be present in Eq.~(\ref{fittingf}). The reduction in the AIC value for the $D=2$ breathing mode is similar, but the extracted non-hydrodynamic mode amplitude is consistent with zero (see Fig.~\ref{fig:two}). Our interpretation of this finding is that there is evidence for the presence of a non-hydrodynamic mode in the $D=2$ quadrupole mode data, while we find no evidence for such a mode in the $D=2$ breathing mode data. The results for the extracted non-hydrodynamic mode damping rates are also shown in Fig.~\ref{fig:two}. Curiously, the extracted damping rate $\Gamma_{Q,nh}$ for $\ln(k_F a)\simeq 1.84,2.8$ is consistent with the kinetic theory analytic result (\ref{eq:bdp}) for an ideal gas in a harmonic trap in kinetic theory using the relation $\omega_\perp \tau_R=0.12 \left(1+\frac{4}{\pi^2}\ln^2(k_F a)\right)$ (cf. Refs.\cite{Brewer2015,2012PhRvA..85a3636B}). 

%\subsection{Three dimensions}

For the case of $D=3$, we analyze raw data by Kinast et al. in Ref.~\cite{2005PhRvL..94q0404K} for the breathing mode oscillations of a gas of $N\simeq2\times 10^{5}$ $^6{\rm Li}$ atoms
at a magnetic field of $B=840$ G (close to unitarity) for several different temperatures $T/T_F$ where $T_F=(3 N \lambda)^{1/3}\omega_\perp$, $\lambda\simeq 0.045$, $\omega_\perp\simeq 2\pi\times 1700$ Hz. For each of the $T/T_F$ values, we extract hydrodynamic mode frequencies and damping rates that are consistent with published values \cite{2005PhRvL..94q0404K,2011Sci...331...58C}.
In Fig.~\ref{fig:two}, amplitudes and extracted damping rates of a putative non-hydrodynamic mode are shown. For a total of 1100 data points, the AIC decreases by 80 units when allowing for a non-hydrodynamic mode in Eq.~(\ref{fittingf}).
Extracted amplitudes and damping rates seem for the most part uncorrelated as a function of $T/T_F$, but it is curious to note that for several values of $T/T_F$, the extracted damping rate is consistent with the kinetic theory result (\ref{eq:bdp}) using the relation $\omega_\perp \tau_R=\frac{45 \pi}{4} \frac{\omega_\perp}{T_F}\frac{T^2}{T_F^2}$ \cite{2008PhRvA..78e3609R}. Our interpretation of these results is that the fits for the $D=3$ breathing mode hint at, but do not provide statistically significant evidence for, the presence of a non-hydrodynamic mode.

\paragraph{\bf Conclusions:}

We have argued that there could be a new class of strongly interacting systems (``SIQFs") sharing similar transport properties beyond the now well-established value of shear viscosity over entropy density. If this is the case, then non-hydrodynamic quasinormal modes in SIQFs, theoretically well-established using the AdS/CFT framework, would imply the presence of non-hydrodynamic modes in all other SIQFs. We have studied collective modes in trapped cold Fermi gases close to unitarity in $D=2,3$ both theoretically and by reanalyzing experimental data. Our analysis hints at the presence of non-hydrodynamic modes in cold Fermi gases, but further experimental work would be needed to corroborate our findings.

We believe that a study of non-hydrodynamic modes in ultracold quantum gases would open a new window into understanding transport properties of these systems and other SIQFs. This could help to found a new theory of ``strongly interacting quantum matter" of importance in many subfields of physics.

\begin{acknowledgments}
\paragraph{\bf Acknowledgements:}
 
This work was supported in part by the Department of Energy, DOE award No. DE-SC0008132. We would like to thank M.~ Koschorreck and C.~ Cao for providing us with the raw time evolution data for the experiments in Refs.~\cite{Vogt:2011np,2013PhRvA..87d3612B,2005PhRvL..94q0404K,2011Sci...331...58C} and R.~Grimm, M.~K\"ohl, K.~Rajagopal and J.~Thomas for useful discussions.

\end{acknowledgments}

\bibliographystyle{apsrev} \bibliography{sqifs}

\end{document}